\newif\ifsingle

\ifsingle
\documentclass[11pt,draftclsnofoot, onecolumn]{IEEEtran}		
\else		
\documentclass[10pt,final, twocolumn]{IEEEtran}
\fi

\usepackage{times}
\usepackage{amsmath,dsfont}
\usepackage{amssymb,amsthm}
\usepackage{epsfig,verbatim}
\usepackage{subfigure}
\usepackage{setspace}
\usepackage{color}
\usepackage{cite}
\usepackage{epstopdf}
\usepackage{graphics}
\usepackage{accents}
\usepackage{acronym}
\usepackage[bookmarks,colorlinks]{hyperref}
\usepackage{booktabs}
\usepackage{mathtools}
\usepackage{algorithm}
\usepackage{algorithmic}
\usepackage{enumitem}
\usepackage{pstool}
\usepackage[linewidth=0.5pt]{mdframed}

\ifsingle

\else

\fi 

\acrodef{adc}[ADC]{Analog-to-Digital Convertor}
\acrodef{dac}[DAC]{digital-to-analog convertor}
\acrodef{cs}[CS]{Compressed Sensing}
\acrodef{dtft}[DTFT]{discrete-time Fourier transform}
\acrodef{dnn}[DNN]{deep neural network} 
\acrodef{csi}[CSI]{channel state information}
\acrodef{map}[MAP]{maximum a-posteriori probability}
\acrodef{snr}[SNR]{signal-to-noise ratio}
\acrodef{sinr}[SINR]{signal-to-interference-and-noise ratio}
\acrodef{bs}[BS]{Base Station} 
\acrodef{iot}[IOT]{Interent of Things}
\acrodef{mimo}[MIMO]{Multiple-Input Multiple-Output}
\acrodef{mse}[MSE]{mean-squared error}
\acrodef{pdf}[PDF]{probability density function}
\acrodef{rv}[RV]{random variable}
\acrodef{fec}[FEC]{forward error correction}
\acrodef{rs}[RS]{Reed-Solomon}
\acrodef{lti}[LTI]{linear time-invariant}
\acrodef{wss}[WSS]{wide-sense stationary}
\acrodef{psd}[PSD]{power spectral density}
\acrodef{ser}[SER]{symbol error rate} 
\acrodef{ber}[BER]{bit error rate} 
\acrodef{isi}[ISI]{intersymbol interference}  
\acrodef{awgn}[AWGN]{additive white Gaussian noise} 
\acrodef{ut}[UTs]{User Terminals} 
\acrodef{mmw}[mmWave]{millimeter wave}
\acrodef{ris}[RIS]{reconfigurable intelligent surface} 
\acrodef{dma}[DMA]{Dynamic Metasurface Antenna} 
\acrodef{5G}{fifth generation}

\IEEEoverridecommandlockouts

\title{ Near-field  Wireless Power Transfer for 6G Internet-of-Everything Mobile Networks: Opportunities and Challenges
}

\author{  
	\IEEEauthorblockN{Haiyang Zhang, Nir Shlezinger, Francesco Guidi, Davide Dardari, Mohammadreza F. Imani, and Yonina C. Eldar\\
	} 

}
 
\begin{document}
	
	\maketitle
 	\pagestyle{empty}  
\thispagestyle{empty} 

\begin{abstract}
 Radiating wireless power transfer (WPT) brings forth the possibility to cost-efficiently charge wireless devices without requiring a wiring infrastructure. As such, it is expected to play a key role in the deployment of limited-battery communicating devices, as part of the 6G enabled Internet-of-Everything (IoE) vision.  To date, radiating WPT technologies are mainly studied and designed assuming that the devices are located in the far-field region of the power radiating antenna, resulting in a relatively low energy transfer efficiency. However, with the transition of 6G systems to mmWave frequencies combined with the usage of large-scale antennas, future WPT devices are likely to operate in the radiating near-field (Fresnel) region. In this article, we provide an overview of the opportunities and challenges which arise from  radiating near-field WPT. 
 In particular, we discuss about the possibility to realize beam focusing in near-field radiating conditions, and  highlight its possible implications for WPT in future {IoE} networks.
 Besides, we overview some of the design challenges and research directions which arise from this emerging paradigm, including its simultaneous operation with wireless communications, radiating waveform considerations, hardware aspects, and  operation with typical antenna architectures.
\end{abstract}

\section{Introduction}

Future wireless communication technologies are expected to pave the way to the Internet-of-Everything (IoE) vision, where a multitude of diverse devices communicate over a wireless media. Many of these devices are portable and can be powered by a limited battery or even be battery-less, giving rise to the need to energize them in a simple and efficient manner. One such technology, which is the focus of growing research attention, is wireless power transfer (WPT) \cite{lu2014wireless}. WPT allows to {power up or} charge wireless devices without requiring a wiring infrastructure.  


Generally speaking, there are three types of techniques to implement WPT: inductive coupling, magnetic resonance coupling, and electromagnetic (EM) radiation \cite{zeng2017communications,CosMas:17}. The first two  are typically highly efficient in terms of energy conversion, but require the distance between the power source and the charging devices to be very small. In particular, these methods belong to the class of near-field non-radiating techniques.
Practically,  this implies  distances on the order of the wavelength which translates, for instance, to at most a few centimeters for signalling using an antenna of diameter not larger than 0.1 meter at  millimeter-wave (mmWave) frequencies. Due to these properties, the first two approaches are considered to be the dominant WPT technologies for applications such as vehicle charging, but they are less applicable for IoE scenarios, e.g., charging remote sensors at long ranges. 

\begin{figure}
		\centering
		\includegraphics[width= 3.2in]{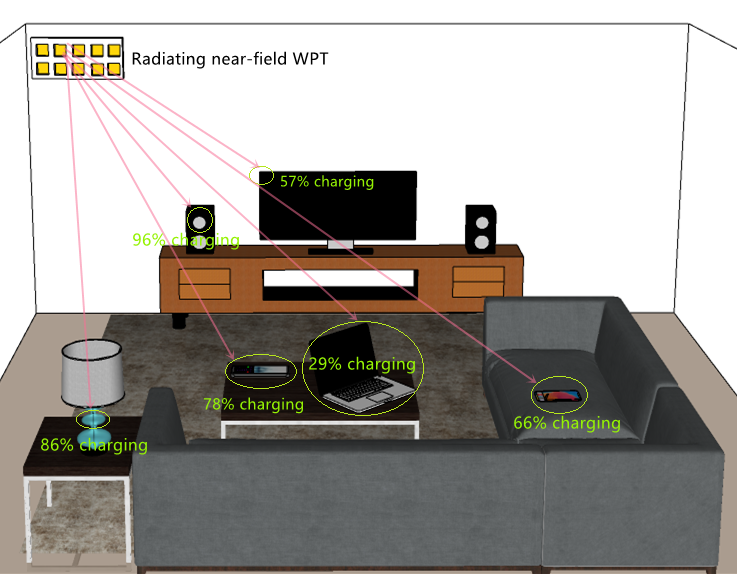}
		\caption{An example of a multi-user WPT system.  }
		\label{fig:model6}
	\end{figure}

The third WPT technique illustrated in Fig.~\ref{fig:model6} utilizes radio frequency (RF) signals to carry energy, allowing to wirelessly power devices over relatively long distances, i.e., similar distances to those over which wireless communication is carried out. This operation has many potential applications for supporting and prolonging the operation of IoE devices in both in-home setups (as depicted in Fig.~\ref{fig:model6}) as well as industrial and commercial settings \cite{CosMas:17}.  
Current research in RF WPT mainly focuses on systems operating in the microwave band (sub 6 GHz) and using relative small antenna arrays so that it is likely that a practical distances (i.e., $>1-2$ meters) the charging devices are located in the far-field propagation regime \cite{CosMas:17}.  In such setups, the radiating wavefront obeys the conventional plane wave model. A key drawback of RF radiation-based WPT in the far-field regime stems form its limited energy transfer efficiency. This is because the beam width from an aperture is limited by diffraction, such that only a fraction of power supplied by the source is captured by the receiver \cite{smith2017analysistransfer}.

	\begin{figure*}
		\centering
		\includegraphics[width= 0.8\linewidth]{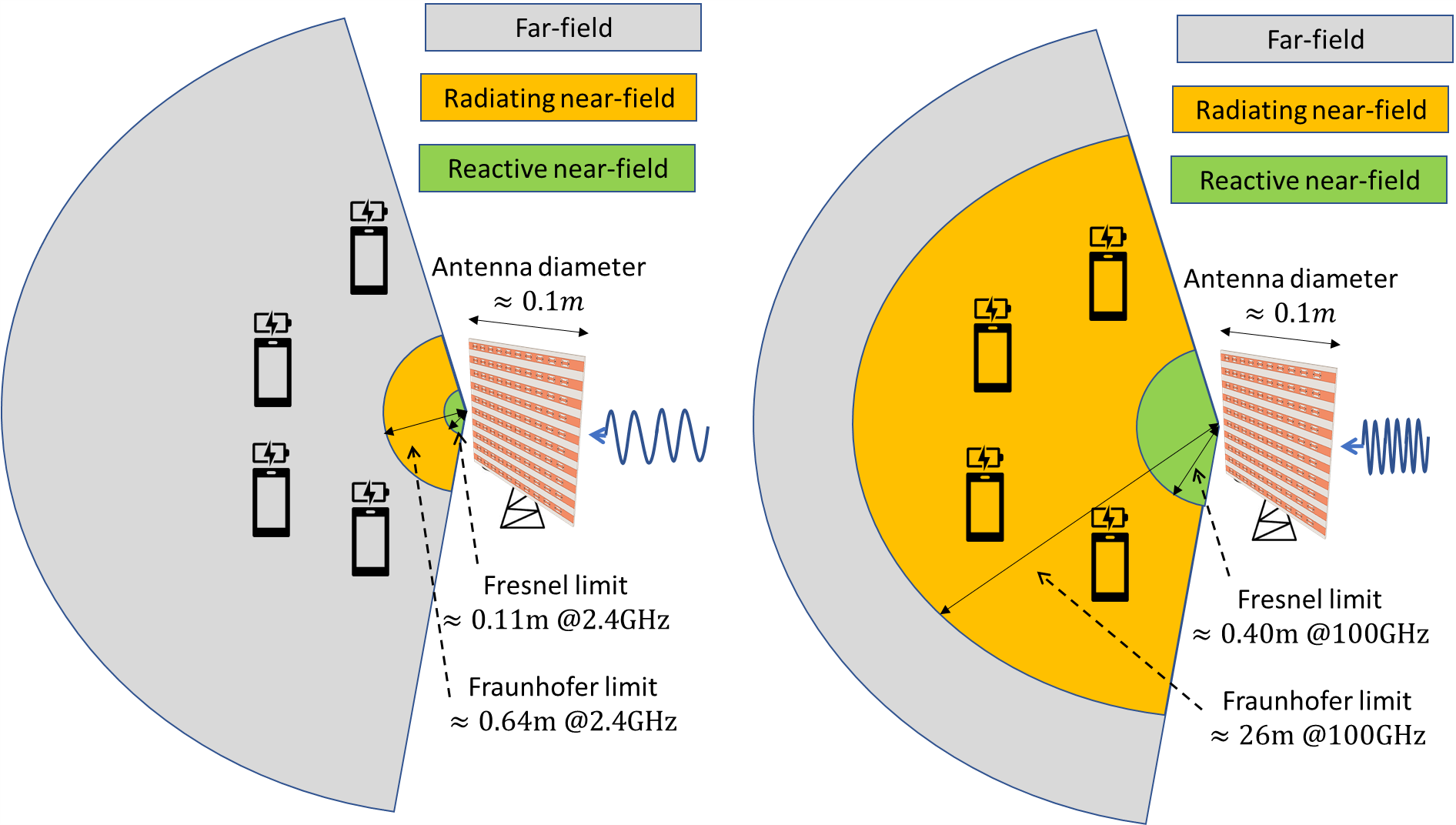}
		\caption{Illustration of the far-field, radiating near-field, and reactive near-field regions when radiating from an antenna with diameter of $0.1$ meters at frequencies of $2.4$ GHz (left) and $100$ GHz (right). This illustration demonstrates that in the mmWave regime, EM-based WPT from a medium-sized antenna often results in the receivers being located in the radiating near field.}
		\label{fig:NearVsFarField1}
	\end{figure*}

6G wireless technologies will utilize large scale antenna arrays to communicate with user devices at the mmWave and sub-THz bands \cite{saad2019vision}. Therefore, 6G-based IoE devices are expected to transmit and receive EM signals while operating in the {\em radiative near-field} \cite{guidi2019radio}.
In such regimes, devices located in distances ranging from a few centimeters to several tens of meters reside in the radiating near-field regime, as illustrated in Fig.~\ref{fig:NearVsFarField1}. Here, one can no longer approximate the spherical wavefornt of the EM field as a plane wave. For wireless communications, operation in the radiative near-field enables the formulation of highly focused beams (as opposed to conventional directional beams achievable in the far field), which can be harnessed to mitigate interference in  multi-user communications \cite{zhang2021beam}. The fact that IoE devices are likely to receive signals in the near-field regime combined with the potential beam focusing capabilities arising in such domains motivate the exploration of radiating near-field WPT for 6G IoE networks.

In this article, we overview opportunities and challenges associated with EM-based WPT in the radiating near-field region. We begin by (i) reviewing the operation of radiating near-field WPT systems, clarifying their relevance for 6G devices; (ii) identifying the physical implications of realizing radiating near-field WPT; and (iii) reviewing existing hardware implementation approaches for the radiating power source. Then, we elaborate on the possibility of exploiting the spherical wavefronts of the EM power signals to implement {\em energy focusing}, identifying its potential advantages compared to far-field WPT.
To demonstrate the potential gains of energy focusing, we  provide a numerical demonstration showing how radiating near-field WPT can simultaneously power multiple wireless devices in an efficient manner with minimal energy pollution. 

Next, we identify key design challenges and the corresponding potential research directions associated with the radiating near-field WPT paradigm. In particular, we focus on the unexplored algorithmic challenges arising from the need to realize energy focusing for 6G IoE networks, including the dependence on accurate channel estimation, adaptation to typical massive antenna array technologies, and dedicated waveform design. In addition, we discuss how near-field WPT can be combined with multi-user wireless information transfer via, e.g., simultaneous wireless information and power transfer (SWIPT) in the near field, and clarify the new design considerations arising from near-field operation.

\section{Radiating Near-field WPT}

\subsection{Radiating WPT Systems}
\begin{figure*}
		\centering
		\includegraphics[width= 1\linewidth]{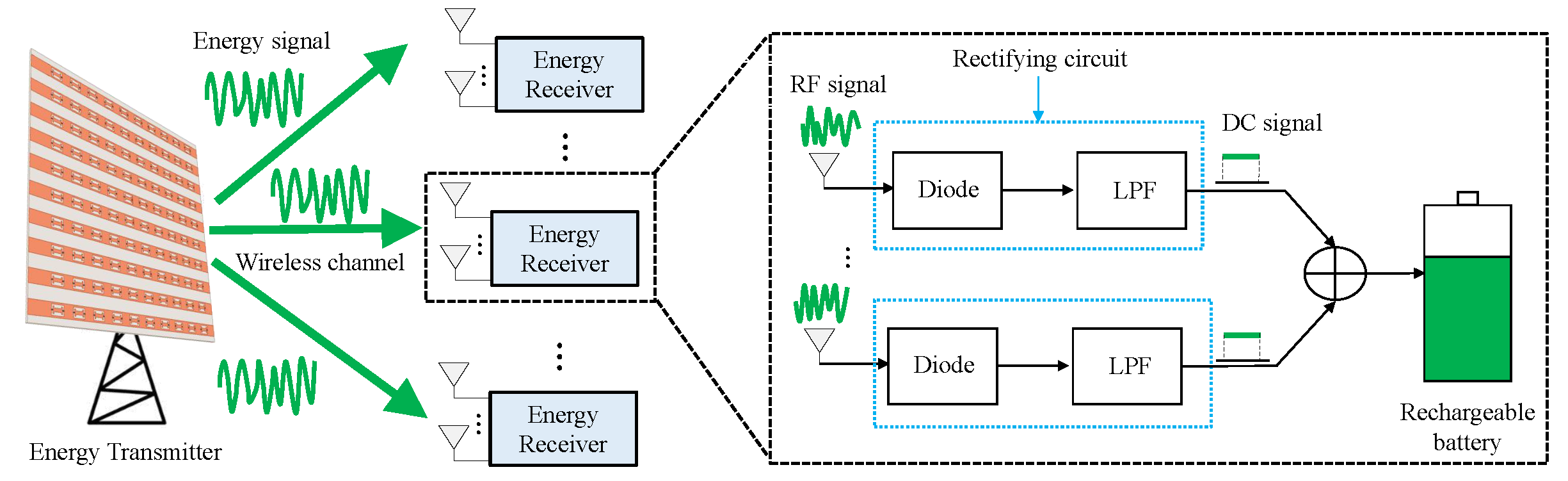}
		\caption{A block diagram of a generic radiating WPT system and the energy receiver structure.}
		\label{fig:WPTBlock}
	\end{figure*}

Radiating WPT systems allow an energy transmitter to charge multiple remote devices, referred to as energy receivers. As illustrated in 
Fig.~\ref{fig:WPTBlock}, in such multi-user radiative WPT setups the energy transmitter typically utilizes an antenna with multiple elements in order to direct the energy towards the receivers, which can also be equipped with  multiple antennas. The energy is transferred by transmitting modulated RF signals to the  energy receivers over a wireless media.  The received RF energy signals are then converted into direct current (DC) signals that 
can be used to charge batteries. In practice, the RF-to-DC conversion is often implemented using a rectifying circuit, which consists of a diode and a low-pass filter (LPF), shown as Fig.~\ref{fig:WPTBlock}. Antennas and RF-to-DC conversion circuits are typically jointly designed and referred to as \emph{rectenna} \cite{CosMas:17}.

Thanks to the broadcast characteristic of wireless channels, the radiative WPT technique is capable of simultaneously charging multiple low-power devices without requiring dedicated wiring. This advantage makes radiative WPT an appealing technique for 6G IoE networks, where a multitude of low-power mobile devices need to be charged frequently. Furthermore, radiative WPT  transmitters can power different mobile energy receivers in a highly controllable manner by designing the transmitted energy signals (including transmit power, time, and frequency). This flexibility enables radiative WPT systems to power low-power devices of 6G IoE systems with different service requirements efficiently. For example, the energy transmitter may charge some  devices when they are idle, and power other devices periodically or on-demand.

In order to charge wireless devices efficiently and safely, there are several requirements on radiating WPT systems one has to account for in practice. The first requirement is to achieve high end-to-end wireless energy transfer efficiency, i.e., have as much of the radiated energy be used for charging the receiver. The energy efficiency is mainly limited by distance-dependent signal attenuation. The distances over which radiated WPT should operate vary between use cases, where a typical distance for indoor settings is on the order of a few meters (e.g., residential rooms) up to several tens of meters (such as halls and corridors).
An additional core requirement is to achieve low energy pollution,  e.g.,  avoid having dominant energy signals in locations where the charging devices are not present. Such energy pollution has a negative effect in terms of human exposure, as well as on communication data transmissions (interference).

\subsection{Physical Implications}
The expected operation of 6G networks, which will utilize large antenna arrays and high-frequency signals, gives rise to several implications for radiating WPT systems in light of the requirements detailed in the previous section. 
It is first noted that high-frequency RF signals are sensitive to obstacles and suffer high signal propagation attenuation. As a result, utilizing high-frequency RF signals in radiating WPT systems may result in reduced reliability to which special care should be given during the design, as it will be highlighted in Section \ref{sec:directions}. On the other hand, the short wavelengths of high-frequency signals facilitates packing a large number of antenna elements on a given area forming a large antenna array of limited physical aperture. The antenna gain stemming from using large antenna arrays bears the potential of compensating for the attenuation of high-frequency signal propagation over distance, resulting in a relatively high energy transfer efficiency. Furthermore, arrays with an extremely large number of antennas will allow to form laser-like beams and thus to generate highly directed energy radiation, which cannot only further increase the energy transfer efficiency, but also achieve low energy pollution. 
As already mentioned, the combination of large antenna arrays with high-frequency implies that WPT devices located in the expected operation distances are likely to reside in the radiating near-field (Fresnel) region. The radiating near-field region is the region between the reactive near-field region and far-field region, i.e., the distance between the energy transmitter and the energy receiver is greater than the Fresnel limit but less than the Fraunhofer limit \cite{BalB:16}. 
For medium-sized antennas on the order of tens of centimeters, combined with signalling at millimeter waves, the expected operation distances of radiating WPT systems, which range from a few to tens of meters, reside between these limits. 

Different from the reactive near-field, where the coupling between antennas is mainly of magnetic nature with a fast decreasing with the distance, 
and the far-field, where plane wave propagation holds, in the radiating near-field the waveform is almost spherical and radiation pattern varies significantly with distance. 
This feature of the radiating near-field implies that some existing results in the radiating WPT literature  derived in the far-field will no longer hold, as explained in Section \ref{sec:directions}. Nonetheless, the expected near-field operation can be exploited to satisfy the requirements for realizing high energy transfer efficiency and low energy pollution. This is achieved using near-field energy focusing, which is discussed and illustrated in Section~\ref{sec:focusing}, using antenna implementations discussed in the sequel.

\subsection{Hardware Implementation}
\label{ssec:hardware}

Typical antenna array transmitters used for implementing 6G communication can be reconfigured to enable wireless transfer of power. The primary distinction in implementing WPT in radiative near-field is that the amplitude and phase of the fields radiated by different antenna elements are configured to constructively interfere at a given focal spot located in the Fresnel region. A receiver placed at this focal spot can thus capture the focused power. In this framework, the antenna element complex amplitude is decided based on the relative location of the antenna element and the desired focal sport, which is distinct from far-field beam steering where the phase profile of the antenna elements is realized to form a plane wave in a desired direction (the distance to the receiver does not matter). At lower mmWave frequencies, the most common technique to realize the desired is to use (active) phased arrays \cite{nepa2017near}. Various dedicated hardware have also been proposed for forming focal spots, including leaky-wave antennas, metasurfaces, or reflectarrays. In the former case, a guided mode is gradually perturbed such that a converging leaky mode is formed which interferes constructively at a focal spot. In the latter case,  metasurface or reflectarray antennas realize a hologram which focuses the signal at a prescribed location. Transferring these ideas to the sub-THz frequencies  envisioned to be utilized in 6G \cite{saad2019vision}, while simple in principle, is a work in progress. In particular, the antenna structure needs to be able to reconfigure its focal spot or form multiple focal spots, while also delivering desired performance for communication purposes. The specific configuration to realize this operation is yet to be decided and is an ongoing research.

The receiver antenna used for WPT in radiative near field also needs to be designed carefully. An important consideration is that the field impinging on it is a converging spherical wave (in contrast to plane waves used in far-field). Antenna arrays as well as perfect absorbing metasurfaces have been proposed and successfully demonstrated for this purpose at lower mmWave frequencies. Another consideration of hardware implementation is the rectifying circuitry. Current works estimate the maximum power conversion efficiency of around $60\%$ at lower mmWave bands. As a result, rectenna hardware that can deliver higher power conversion efficiencies at mmWave and sub-THZ is still a work in progress.
 
\section{Energy Focusing}
\label{sec:focusing}

\subsection{Near-Field Energy Focusing}
In conventional far-field operation, one can utilize antenna arrays to generate directed RF signalling via {\em beam steering}. Roughly speaking, the transmitted energy is broadcast within a given angular sector whose width decreases with the size of the antenna array. For a given beam width, as long as the receiver goes far from the array, the WPT efficiency decreases because it intercepts a smaller angle of the beam. Therefore, WPT operating in far-field condition might be very inefficient as most of the energy is cast  away. Instead, when operating in the radiating near-field, the non-negligible spherical wavefront of RF signals enables generating beams which are not only steered in a given direction, but are actually focused on a specific desired location \cite{nepa2017near}. This capability is referred to as {\em beam focusing}.  
The ability to generate focused beams can be exploited by wireless communications systems to facilitate radio positioning \cite{guidi2019radio}. Furthermore, it was recently shown in \cite{zhang2021beam} that beam focusing can facilitate downlink transmissions, where it is exploited to mitigate interference and allow reliable communications with different users lying at the same relative angle. In the context of radiating near-field WPT, where the EM signals are used also to transfer energy, the ability to direct the signal intended for each user at a specific location results in {\em energy focusing}.

Energy focusing brings forth several core advantages to radiating near-field WPT systems. 
First, it enhances the energy transfer efficiency compared to directive radiation in the far-field; the fact that the power transmitter can focus its radiated energy on the exact locations where the charging devices results in more energy being received. Furthermore, it reduces energy pollution and limits human exposure to radiated energy. In fact, one can envision IoE devices being simultaneously utilized by a human operator and charged via radiating near-field WPT, while the energy waves are focused only on the charging device such that the human is hardly exposed to its energetic radiation. This capability is expected to notably facilitate the charging of 6G IoE devices in indoor settings.

\begin{figure} 
  \centering 
    \subfigure[Single energy receiver located at the near-field region]{ 
    \label{fig:near-field}
    \includegraphics[width=3.6in]{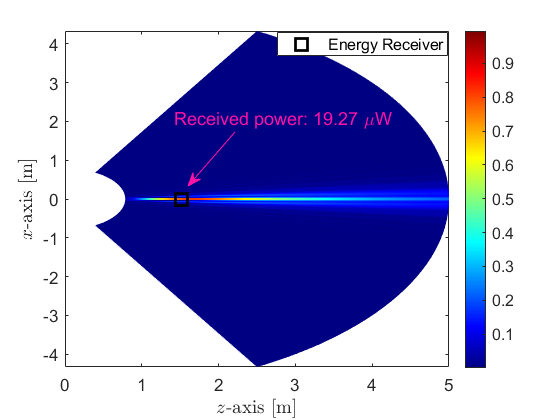} 
  } 
  \subfigure[Single energy receiver located at the far-field region]{ 
    \label{fig:far-field}
    \includegraphics[width=3.6in]{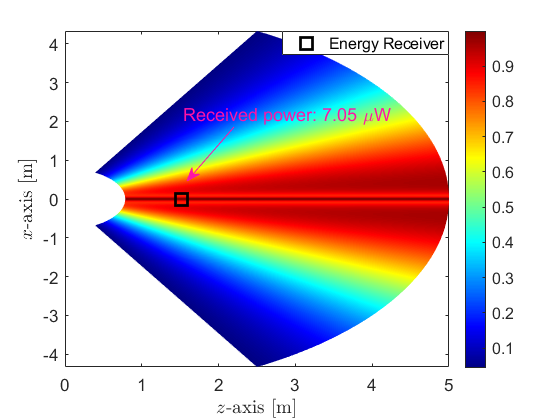} 
  } 
   \caption{The normalized received power of the energy receiver located at the: (a) near-field region; (b) far-field region.}    \label{fig:single_user} 
\end{figure}

\begin{figure} 
  \centering 
    \includegraphics[width=3.6in]{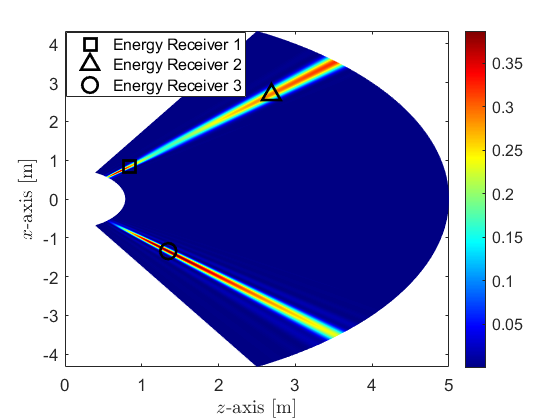} 
   \caption{The normalized received power of multiple energy receivers located in the near-field region.}   
   \label{fig:Multi_user} 
\end{figure}

\subsection{Numerical Results}
\label{ssec:Sim}
To demonstrate the potential of near-field energy beam for radiating WPT, we next present some representative numerical results. To that aim, we consider a near-field WPT system where the energy transmitter is equipped with a fully-digital planar array positioned in the $xy$-plane, and the single-antenna energy receivers are positioned in the $xz$-plane. The antenna size is $30~ {\rm cm} \times 30~ {\rm cm}$ with a half wavelength separation between each element. The maximum transmit power is set to be $1~$W,  
while the RF-to-DC energy conversion efficiency at the receivers is $0.5$. The wireless channels are generated according to the near-field wireless channel model detailed in \cite{zhang2021beam}.
In Figs. \ref{fig:single_user} and \ref{fig:Multi_user}, we depict the normalized received power at the energy receivers,  defined as the ratio of the received power of an energy receiver to its corresponding channel gain.

Fig.~\ref{fig:single_user} depicts the numerically evaluated normalized received power at each point of the predefined region in the $xz$-plane. The energy transmission scheme is optimized to maximize the received power of the targeted energy receiver. For Fig. \ref{fig:single_user}(a), the carrier 
frequency is $28~$GHz, resulting in the target energy receiver being located in the near-field region, while for Fig. \ref{fig:single_user}(b), the carrier 
frequency is $1.2~$GHz and the target energy receiver is located in the far-field region.
It is from Fig. \ref{fig:single_user}(a) that in the near-field case, by virtue of a narrow energy beam, most of the transmitted energy can be focused around the target energy receiver area, and thus harvested by the energy receiver with extremely high efficiency.
By contrast, for the far-field case as shown in Fig. \ref{fig:single_user}(b), energy can only be transmitted towards a direction with a comparatively wider energy beam. Consequently, far-field signalling results in the target energy receiver harvesting only $36\%$ of the energy obtained in the near-field.  
This gain is achieved despite the fact that the system in Fig. \ref{fig:single_user}(b) operates at a lower frequency and hence with a lower isotropic path loss.
Additionally, the comparison between Figs. \ref{fig:single_user}(a) and \ref{fig:single_user}(b)  demonstrates that for the  radiating near-field WPT system, energy beam focusing does not only enhance the energy transfer efficiency, but also reduces energy pollution. 


Fig. \ref{fig:Multi_user} shows the normalized received energy of multiple energy receivers where the locations of receivers are randomly generated within the radiating near-field region. The transmitted energy beam vector is the sum of three beam vectors, each designed to maximize the harvested energy of their respective target energy receiver. 
From Fig. \ref{fig:Multi_user} it os observed that the closer the energy receiver is to the transmitter, the better the focusing performance will be, i.e., the energy focusing region is much smaller for energy receiver 1 than energy receiver 2 and energy receiver 3. Moreover, it is also observed that energy beam focusing is capable of flexibly serving multiple energy receivers with minor energy pollution even if they are not distinguishable in the angular direction by the transmitter. For instance, as the figure shows, although energy receivers $1$ and $2$ lie in the same angular direction, energy focusing regions can be generated separately for each of them. Such a distinguishing capability of the near-field energy beam focusing enables its potential use in 6G IoE networks, where ultra high devices densities are expected even in the Fresnel region.

\section{Design challenges and research directions}
\label{sec:directions}

As stated above, radiating near-field WPT enables high energy transfer efficiency by using energy beam focusing, making it an appealing technique for wirelessly charging massive low-power devices of future IoE scenarios. However, approaching the performance gain of radiating near-field WPT also requires to carefully address several challenges, giving rise to many interesting new research opportunities. In the following, we briefly discuss some of these potential challenges/opportunities. 

\subsection{Channel Estimation} 
The performance gain of energy beam focusing depends highly on the accuracy of channel state information (CSI). Thus, the CSI of near-field wireless channels needs to be accurately estimated, which in fact is a great challenge in practice, especially when using large antenna arrays. On the other hand, when transmitting energy signals with high frequencies such as mmWave bands,  wireless channels are basically dominated by line-of-sight (LoS) channel models. As a result, the CSI will be highly related to the relative locations between the energy transmitter and receivers. Therefore, joint localization and channel estimation for radiating near-field WPT systems is an aspect worthy of further investigation.

\subsection{Simultaneous WPT and Information Transfer}

RF signals, apart from delivering energy, are typically used to carry information in wireless communications. Therefore, in recent years, a paradigm combining wireless power transfer and wireless communications, coined SWIPT \cite{xu2014multiuser}, has been extensively investigated in the far-field. In SWIPT systems, the dual functions of WPT and information delivery are jointly designed on a common hardware platform, achieving simultaneously high-frequency efficiency and  low hardware cost. When SWIPT systems operate in the radiating near-field region, where a spherical wave is used instead of the typical far-field plane wave,
some known conclusions or results derived in the far-field setting may not be valid any more. 
As a consequence, it is necessary to rethink SWIPT systems in the radiating near-field case. Some  potential research topics are:
\begin{itemize}
\item For multi-user SWIPT systems, where an energy transmitter communicates and powers multiple remote devices, the ability to form focused beams in the near-field brings forth some exciting opportunities. For instance, one can transfer both information and power to multiple devices lying in the same angular direction with minimal cross-interference and while hardly radiating energy towards any location other than that where the devices reside. 

\item 
In setups where the information receivers and energy receivers are separate, near-field operation gives rise to new design considerations. For once, dedicated energy beams are known not to be necessary in the far-field case~\cite{xu2014multiuser}. This is because the comparatively wide information beams can also be used to charge energy receivers, and meanwhile, additional energy beams may cause co-channel interference to the information receivers. 
However, for the same system setup within the radiating near-field region, dedicated energy beams are required. This can be explained by the fact that information beams mainly focus on the locations of information receivers, and thus the energy receivers may not harvest enough energy if no energy beams are sent. 
Fortunately, thanks to the capability of energy focusing, energy beams can be designed  not to cause severe interference to information receivers. 
Moreover, as beam focusing not only enhances the energy transfer efficiency, but also increases the achievable rate of wireless communication systems \cite{zhang2021beam}, it is expected that beam focusing can achieve a larger harvested energy-rate region than that achieved by far-field beam steering. Thus, it is  desirable to design transmission schemes for radiating near-field SWIPT systems.

\item For SWIPT systems with information security constraints, where the information messages sent to the information receivers should be kept secret to energy receivers who are regarded as potential eavesdroppers \cite{liu2014secrecy}, radiating near-field beam focusing offers a potential way to enhance the achievable secrecy rate of information receivers. In this case, the nature of beam focusing is utilized to focus information signals on the locations of information receivers while reduces information leakage to potential eavesdroppers. 

\item The combination of WPT with wireless information transfer can also be envisioned to be co-designed with alternative usages of near-field RF signals, such as localization and sensing. Moreover, the future integration of 6G systems with Radio Frequency Identification (RFID) and Real-Time Location Systems (RTLS) gives rise to new possibilities involving near-field WPT, where for instance one can power up, localize, communicate with the surrounding passive tags using a mobile device or  an access point providing WPT \cite{DarDecGueGui:C16}.

 \end{itemize}

\subsection{Antenna Architectures}
The ability to generate energy focusing is largely dependent on the antenna architecture used by the energy transmitter. 
In practice, the larger the antenna array is, the smaller the beam waist will be, yielding improved beam focusing performance. Meanwhile, a large antenna array means a large radiating near-field region for a fixed element spacing. Therefore, a large antenna array is highly desirable to enhance the energy transfer efficiency and enlarge the radiating near-field region. 

In Subsection~\ref{ssec:Sim} we demonstrated the gains of energy focusing while using a fully-digital antenna array, in which each element is fed using a dedicated RF chain. Implementing such large-scale  fully-digital antenna arrays may be too costly in many applications due to the excessive hardware and its effect on cost, power consumption, and physical size. As discussed in Subsection~\ref{ssec:hardware}, hardware implementations of feasible antenna architectures for such operation is an area of active research, which is expected to have a notable impact on the successful deployment of radiative WPT systems. 

Another important direction of investigation is how to make WPT robust to channel blockage impairments typical at mmWave. This is especially relevant for 6G IoE devices applied in urban and indoor settings. A recent technology is given by intelligent reflecting metasurfaces which can be exploited not only for communication but also to extend the coverage of WPT systems with limited complexity \cite{wu2019towards}.
Finally, implementing energy receiver architectures with high conversion efficiency is an important need to be addressed for future WPT systems. This includes both RF-to-DC circuits such as rectennas,  as well as receive antennas supporting SWIPT operation.  

\subsection{Beam and Waveform Design} 
In the previous sections, we discussed beam focusing design for maximizing the energy transfer efficiency based on the linear energy harvesting model assumption, where the rectenna implementing RF-to-DC conversion is independent of its input waveform. 
Nonetheless, there are still several key algorithmic aspects of near-field energy transmission design which should be addressed. For once, the ability to limit energy pollution motives the design of intelligent algorithms that are {\it ambient aware}, allowing to reduce the waste of energy at most and to account for locations where radiation is not wanted. Furthermore, the incorporation of novel antenna architectures with analog precoding capabilities brings forth design challenges in generating focused beams \cite{zhang2021beam}. 

In addition, the RF-to-DC conversion efficiency of the rectenna also depends on the input waveform when its input RF signals' power is large \cite{clerckx2016waveform}. 
Waveform design highly relies on wireless channels. As the channel propagation model of the radiative near-field is essentially different from the far-field one, the transmit waveform design (including energy beam focusing) for radiative near-field WPT systems is also an important research topic, especially if the optimization has to be performed jointly with communication, localization and sensing.


\section{Conclusion}
To date, radiating WPT is mainly studied and designed in the far-field region, resulting in relatively low energy transfer efficiency. However, with the transition of 6G systems to high-frequency combined with the usage of large-scale antennas, WPT devices may easily operate in the radiating near-field (Fresnel) region, where the conventional plane wave propagation assumption in far-field is no longer valid. In this article, we provided an overview of the opportunities and challenges of radiating near-field WPT systems. In particular, we first discussed the key characteristics of near-field radiation where the spherical waveform propagation model holds, and clarified its implications on  WPT. Then, we presented the technique of energy beam focusing, and highlighted its advantages for IoE networks. We concluded with some of the design challenges and potential research directions, which are expected to pave the way for implementing near-field radiating WPT systems for wirelessly charging the devices of future 6G IoE networks.

	\bibliographystyle{IEEEtran}
	\bibliography{IEEEabrv,refs}

\begin{thebibliography}{10}
\providecommand{\url}[1]{#1}
\csname url@samestyle\endcsname
\providecommand{\newblock}{\relax}
\providecommand{\bibinfo}[2]{#2}
\providecommand{\BIBentrySTDinterwordspacing}{\spaceskip=0pt\relax}
\providecommand{\BIBentryALTinterwordstretchfactor}{4}
\providecommand{\BIBentryALTinterwordspacing}{\spaceskip=\fontdimen2\font plus
\BIBentryALTinterwordstretchfactor\fontdimen3\font minus
  \fontdimen4\font\relax}
\providecommand{\BIBforeignlanguage}[2]{{%
\expandafter\ifx\csname l@#1\endcsname\relax
\typeout{** WARNING: IEEEtran.bst: No hyphenation pattern has been}%
\typeout{** loaded for the language `#1'. Using the pattern for}%
\typeout{** the default language instead.}%
\else
\language=\csname l@#1\endcsname
\fi
#2}}
\providecommand{\BIBdecl}{\relax}
\BIBdecl

\bibitem{lu2014wireless}
X.~Lu, P.~Wang, D.~Niyato, D.~I. Kim, and Z.~Han, ``Wireless networks with {RF}
  energy harvesting: A contemporary survey,'' \emph{{IEEE} Commun. Surveys
  Tuts.}, vol.~17, no.~2, pp. 757--789, 2014.

\bibitem{zeng2017communications}
Y.~Zeng, B.~Clerckx, and R.~Zhang, ``Communications and signals design for
  wireless power transmission,'' \emph{{IEEE} Trans. Commun.}, vol.~65, no.~5,
  pp. 2264--2290, 2017.

\bibitem{CosMas:17}
A.~Costanzo and D.~Masotti, ``Energizing {5G}: Near- and far-field wireless
  energy and data transfer as an enabling technology for the {5G IoT},''
  \emph{{IEEE} Microw. Mag.}, vol.~18, no.~3, pp. 125--136, 2017.

\bibitem{smith2017analysistransfer}
D.~R. Smith, V.~R. Gowda, O.~Yurduseven, S.~Larouche, G.~Lipworth, Y.~Urzhumov,
  and M.~S. Reynolds, ``An analysis of beamed wireless power transfer in the
  fresnel zone using a dynamic, metasurface aperture,'' \emph{Journal of
  Applied Physics}, vol. 121, no.~1, p. 014901, 2017.

\bibitem{saad2019vision}
W.~Saad, M.~Bennis, and M.~Chen, ``A vision of 6{G} wireless systems:
  Applications, trends, technologies, and open research problems,''
  \emph{{IEEE} Netw.}, vol.~34, no.~3, pp. 134--142, 2019.

\bibitem{guidi2019radio}
F.~Guidi and D.~Dardari, ``Radio positioning with {EM} processing of the
  spherical wavefront,'' \emph{{IEEE} Trans. Wireless Commun.}, vol.~20, no.~6,
  pp. 3571--3586, 2021.

\bibitem{zhang2021beam}
H.~Zhang, N.~Shlezinger, F.~Guidi, D.~Dardari, M.~F. Imani, and Y.~C. Eldar,
  ``Beam focusing for near-field multi-user {MIMO} communications,''
  \emph{arXiv preprint arXiv:2105.13087}, 2021.

\bibitem{BalB:16}
C.~A. Balanis, \emph{Antenna Theory: analysis and design}.\hskip 1em plus 0.5em
  minus 0.4em\relax New Jersey, USA: Wiley, 2016.

\bibitem{nepa2017near}
P.~Nepa and A.~Buffi, ``Near-field-focused microwave antennas: Near-field
  shaping and implementation.'' \emph{{IEEE} Antennas Propag. Mag.}, vol.~59,
  no.~3, pp. 42--53, 2017.

\bibitem{xu2014multiuser}
J.~Xu, L.~Liu, and R.~Zhang, ``Multiuser {MISO} beamforming for simultaneous
  wireless information and power transfer,'' \emph{{IEEE} Trans. Signal
  Process.}, vol.~62, no.~18, pp. 4798--4810, 2014.

\bibitem{liu2014secrecy}
L.~Liu, R.~Zhang, and K.-C. Chua, ``Secrecy wireless information and power
  transfer with {MISO} beamforming,'' \emph{{IEEE} Trans. Signal Process.},
  vol.~62, no.~7, pp. 1850--1863, 2014.

\bibitem{DarDecGueGui:C16}
D.~Dardari, N.~Decarli, A.~Guerra, and F.~Guidi, ``The future of
  {Ultra-Wideband} localization in {RFID},'' in \emph{2016 IEEE International
  Conference on RFID (RFID) (IEEE RFID 2016)}, Orlando, USA, May 2016.

\bibitem{wu2019towards}
Q.~Wu and R.~Zhang, ``Towards smart and reconfigurable environment: Intelligent
  reflecting surface aided wireless network,'' \emph{{IEEE} Commun. Mag.},
  vol.~58, no.~1, pp. 106--112, 2019.

\bibitem{clerckx2016waveform}
B.~Clerckx and E.~Bayguzina, ``Waveform design for wireless power transfer,''
  \emph{{IEEE} Trans. Signal Process.}, vol.~64, no.~23, pp. 6313--6328, 2016.

\end{thebibliography}

\begin{IEEEbiographynophoto}{Haiyang Zhang}
 is a postdoctoral researcher at Weizmann Institute of Science, Israel.
\end{IEEEbiographynophoto}	
\vskip -2\baselineskip plus -1fil

\begin{IEEEbiographynophoto}{Nir Shlezinger}  is an Assistant Professor in the School of Electrical and Computer Engineering in Ben-Gurion University, Israel.
\end{IEEEbiographynophoto}
\vskip -2\baselineskip plus -1fil

\begin{IEEEbiographynophoto}{Francesco Guidi} is a Researcher at the National Research Council of Italy, Institute of Electronics, Computer and Telecommunication Engineering, Italy. 
\end{IEEEbiographynophoto}
\vskip -2\baselineskip plus -1fill

\begin{IEEEbiographynophoto}{Davide Dardari} is a Full Professor at the  Department of Electrical, Electronic, and Information Engineering “Guglielmo Marconi” - DEI-CNIT, University of Bologna, Italy.
\end{IEEEbiographynophoto}	
\vskip -2\baselineskip plus -1fill

\begin{IEEEbiographynophoto}{Mohammadreza F. Imani}
is an Assistant Professor in the ECEE school, Arizona State University, USA.  
\end{IEEEbiographynophoto}
\vskip -2\baselineskip plus -1fill

\begin{IEEEbiographynophoto}{Yonina C. Eldar}
is a Professor in the Department of Math and Computer Science, Weizmann Institute of Science, Israel.
\end{IEEEbiographynophoto}	
	
\end{document}